**Is formal QED mathematically consistent?**

by

Dan Solomon
Rauland-Borg Corporation
3450 W. Oakton
Skokie, IL 60077
USA

Email: **dan.solomon@rauland.com**

PAC 11.10-z
(Sept 5, 2005)

**<u>Abstract</u>**

The formal structure of quantum electrodynamics consists of various elements. These include the Schrödinger equation which evolves the system forward in time, the vacuum state which is assumed to be the state with a free field energy of zero, and the principle of gauge invariance. In this paper we will examine the mathematical consistency of these elements. It will be shown that these elements, which are basic to the formal structure of the theory, are not mathematically consistent.

## **1. Introduction**

In this paper we will examine the formal structure of quantum electrodynamics (QED). We will consider a quantized fermion field which is acted on by an applied classical electromagnetic field. QED is a mathematical theory in that it consists of various elements that are assumed to be mathematically consistent. These elements include the following: First, there is the Schrödinger equation which governs the dynamics of the theory, that is, it evolves the system forward in time. Second, there is the vacuum state $|0\rangle$. The key feature of the vacuum state is that its free field energy is assumed to be zero and all other states have a free field energy greater than zero. The free field energy is the expectation value of the free field Hamiltonian and will be formally defined later. A third element of the theory is that it is assumed to be gauge invariant. The question that will be addressed here is whether or not these elements are mathematically consistent. It will be shown that they are not. Throughout this discussion we assume $\hbar = c = 1$.

In the Schrödinger representation of QED the field operators are time independent and the state vector $|\Omega(t)\rangle$ evolves in time according to the Schrödinger equation [1,2],

$$i\frac{\partial}{\partial t}|\Omega(t)\rangle = \hat{H}|\Omega(t)\rangle; \quad -i\frac{\partial}{\partial t}\langle\Omega(t)| = \langle\Omega(t)|\hat{H} \tag{1}$$

where $\hat{H}$ is the Dirac Hamiltonian and is given by,

$$\hat{H} = \hat{H}_0 - \int \hat{\mathbf{J}}(\mathbf{x}) \cdot \mathbf{A}(\mathbf{x},t)d\mathbf{x} + \int \hat{\rho}(\mathbf{x}) \cdot A_0(\mathbf{x},t)d\mathbf{x} \tag{2}$$

where $\mathbf{A}(\mathbf{x},t)$ and $A_0(\mathbf{x},t)$ are the vector and scalar potential, respectively, of the applied classical field, $\hat{\mathbf{J}}(\mathbf{x})$ and $\hat{\rho}(\mathbf{x})$ are the current and change operators, respectively, and $\hat{H}_0$ is the free field Hamiltonian. This is just the Hamiltonian when the electric potential is zero. The current and charge operators are defined in terms of the time independent field operator $\psi(\mathbf{x})$ by,

$$\hat{\mathbf{J}}(\mathbf{x}) = \frac{e}{2}\Big[\psi^\dagger(\mathbf{x}), \boldsymbol{\alpha}\psi(\mathbf{x})\Big] \text{ and } \hat{\rho}(\mathbf{x}) = \frac{e}{2}\Big[\psi^\dagger(\mathbf{x}), \psi(\mathbf{x})\Big] \tag{3}$$

where 'e' is the electric charge.

Assume that $\big|\Omega(t)\big\rangle$ is a normalized state vector. In this case the expectation value of the current and charge operators are defined by,

$$\mathbf{J}_e(\mathbf{x},t) = \big\langle\Omega(t)\big|\hat{\mathbf{J}}(\mathbf{x})\big|\Omega(t)\big\rangle \text{ and } \rho_e(\mathbf{x},t) = \big\langle\Omega(t)\big|\hat{\rho}(\mathbf{x})\big|\Omega(t)\big\rangle \tag{4}$$

and the free field energy is defined by $\big\langle\Omega(t)\big|\hat{H}_0\big|\Omega(t)\big\rangle$.

Now examine the quantity $\xi(t_i \to t_f)$ which is defined by,

$$\xi(t_i \to t_f) = \int_{t_i}^{t_f} dt \int \mathbf{J}_e(\mathbf{x},t)\cdot\mathbf{E}(\mathbf{x},t)d\mathbf{x} \tag{5}$$

where $t_f > t_i$ and where $\mathbf{E}(\mathbf{x},t)$ is the applied electric field and is given in terms of the vector and scalar potential by,

$$\mathbf{E}(\mathbf{x},t) = -\left(\frac{\partial\mathbf{A}(\mathbf{x},t)}{\partial t} + \nabla A_0(\mathbf{x},t)\right) \tag{6}$$

Eq. (5) can also be written as,

$$\frac{d\xi(t_i \to t)}{dt} = \int \mathbf{J}_e(\mathbf{x},t)\cdot\mathbf{E}(\mathbf{x},t)d\mathbf{x} \tag{7}$$

In classical mechanics this would be the rate of increase in the kinetic energy of a charged particle when acted on by an electric field. Now use (6) in (7) to obtain,

$$\frac{d\xi\left(t_i \to t\right)}{dt} = -\int \mathbf{J}_e \cdot \left(\frac{\partial \mathbf{A}}{\partial t} + \nabla A_0\right) d\mathbf{x} = -\frac{\partial}{\partial t}\left(\int \mathbf{J}_e \cdot \mathbf{A} d\mathbf{x}\right) + \int \frac{\partial \mathbf{J}_e}{\partial t} \cdot \mathbf{A} d\mathbf{x} + \int A_0 \nabla \cdot \mathbf{J}_e d\mathbf{x} \quad (8)$$

where integration by parts has been used to obtain the last term in the above expression. Next use (4) and (1) to obtain,

$$\frac{\partial \mathbf{J}_e}{\partial t} = \frac{\partial \langle \Omega(t)|}{\partial t} \hat{\mathbf{J}}(\mathbf{x}) |\Omega(t)\rangle + \langle \Omega(t)| \hat{\mathbf{J}}(\mathbf{x}) \frac{\partial |\Omega(t)\rangle}{\partial t} = i \langle \Omega(t)| \left[\hat{H}, \hat{\mathbf{J}}(\mathbf{x})\right] |\Omega(t)\rangle \quad (9)$$

Use (2) and the fact that the classical quantities $A_0$ and $\mathbf{A}$ commute with $\hat{H}$ in the above to yield,

$$\begin{aligned}
\int \frac{\partial \mathbf{J}_e}{\partial t} \cdot \mathbf{A} d\mathbf{x} &= i \langle \Omega(t)| \left[\hat{H}, \int \hat{\mathbf{J}} \cdot \mathbf{A} d\mathbf{x}\right] |\Omega(t)\rangle \\
&= i \langle \Omega(t)| \left[\hat{H}, \hat{H}_0 + \int \hat{\rho} \cdot A_0 d\mathbf{x} - \hat{H}\right] |\Omega(t)\rangle \\
&= i \langle \Omega(t)| \left[\hat{H}, \hat{H}_0\right] |\Omega(t)\rangle + i \langle \Omega(t)| \left[\hat{H}, \int \hat{\rho} \cdot A_0 d\mathbf{x}\right] |\Omega(t)\rangle
\end{aligned} \quad (10)$$

Next use (1) and (4) to yield,

$$\int \frac{\partial \mathbf{J}_e}{\partial t} \cdot \mathbf{A} d\mathbf{x} = \frac{\partial}{\partial t} \langle \Omega(t)| \hat{H}_0 |\Omega(t)\rangle + \int \frac{\partial \rho_e}{\partial t} \cdot A_0 d\mathbf{x} \quad (11)$$

Use this in (8) to obtain,

$$\frac{d\xi\left(t_i \to t\right)}{dt} = -\frac{\partial}{\partial t}\left(\int \mathbf{J}_e \cdot \mathbf{A} d\mathbf{x}\right) + \frac{\partial}{\partial t} \langle \Omega(t)| \hat{H}_0 |\Omega(t)\rangle + \int A_0 \left(\frac{\partial \rho_e}{\partial t} + \nabla \cdot \mathbf{J}_e\right) d\mathbf{x} \quad (12)$$

Re-arrange terms and integrate from some initial time $t_i$ to some final time $t_f$ to obtain the following expression,

$$\left\langle \Omega\left(t_f\right) \middle| \hat{H}_0 \middle| \Omega\left(t_f\right) \right\rangle = \xi\left(t_i \to t\right)$$
$$+ \int d\mathbf{x}\left(\mathbf{J}_e\left(\mathbf{x}, t_f\right) \cdot \mathbf{A}\left(\mathbf{x}, t_f\right) - \mathbf{J}_e\left(\mathbf{x}, t_i\right) \cdot \mathbf{A}\left(\mathbf{x}, t_i\right)\right) \quad (13)$$
$$- \int_{t_i}^{t_f} dt \int A_0 \left(\frac{\partial \rho_e}{\partial t} + \nabla \cdot \mathbf{J}_e\right) d\mathbf{x} + \left\langle \Omega\left(t_i\right) \middle| \hat{H}_0 \middle| \Omega\left(t_i\right) \right\rangle$$

## 2. The vacuum state

The state vector $\left|0\right\rangle$ is defined as the vacuum state and is the state in which no particles or anti-particles exist. It is an eigenvector of the free field Hamiltonian operator $\hat{H}_0$ with an eigenvalue of zero,

$$\hat{H}_0 \left|0\right\rangle = 0 \quad (14)$$

A complete set of orthonormal eigenstates $\left|n\right\rangle$ can be produced by acting on the vacuum state $\left|0\right\rangle$ by creation operators which produce additional states. The eigenstates $\left|n\right\rangle$ satisfy,

$$\left\langle n \middle| m \right\rangle = \delta_{mn} \quad (15)$$

and

$$\hat{H}_0 \left|n\right\rangle = E_n \left|n\right\rangle \text{ where } E_n > 0 \text{ for } \left|n\right\rangle \neq \left|0\right\rangle \quad (16)$$

Any arbitrary state $\left|\Omega\right\rangle$ can be expresses as a Fourier sum of the eigenstates $\left|n\right\rangle$ (including the vacuum state $\left|0\right\rangle$), e.g.,

$$\left|\Omega\right\rangle = \sum_j c_j \left|n_j\right\rangle \text{ and } \left\langle\Omega\right| = \sum_j c_j^* \left\langle n_j\right| \quad (17)$$

where $c_j$ are the expansion coefficients. Use (15) and (16) to show that,

$$\left\langle \Omega \middle| \hat{H}_0 \middle| \Omega \right\rangle = \sum_j \left|c_j\right|^2 E_j \quad (18)$$

Due to the fact that all the $E_j \geq 0$ it is evident that for an arbitrary state $|\Omega\rangle$ the following relationship holds,

$$\langle\Omega|\hat{H}_0|\Omega\rangle \geq \langle 0|\hat{H}_0|0\rangle = 0 \qquad (19)$$

Therefore the free field energy of any arbitrary state is non-negative.

### 3. Gauge invariance

In Section 1 and 2 we have introduced some elements of the formal theory. In this section we will add another element. We will assume that the theory is gauge invariant. We will try to determine the effect, if any, that this assumption has on the elements of the theory introduced so far. What we are looking for is mathematical consistency. Is the assumption of gauge invariance mathematically consistent with the material presented in the last two sections?

Gauge invariance is an important requirement that quantum theory must satisfy [3]. A change in the gauge is a change in the electric potential that leaves the electric and magnetic fields unchanged. Such a transformation is given by,

$$\mathbf{A} \to \mathbf{A}' - \nabla\chi; \ A_0 \to A_0' + \frac{\partial\chi}{\partial t} \qquad (20)$$

where $\chi(\mathbf{x}, t)$ is an arbitrary real valued function.

If a theory is gauge invariant then the observables of the theory are invariant under a gauge transformation. This includes the current and charge expectation values and the quantity $\xi(t_i \to t_f)$. Now consider the following problem. Assume that at the initial time $t_i$ the electric potential is zero and the state vector is $|\Omega(t_i)\rangle$. Next apply a non-zero electric potential for $t > t_i$, i.e.,

$$\left(\mathbf{A}\left(\mathbf{x},t\right),A_0\left(\mathbf{x},t\right)\right)=0 \text{ at } t=t_i; \quad \left(\mathbf{A}\left(\mathbf{x},t\right),A_0\left(\mathbf{x},t\right)\right)\neq 0 \text{ for } t>t_i \tag{21}$$

Now, under the action of the electric potential, the state vector evolves from the initial state $\left|\Omega\left(t_i\right)\right\rangle$ into some final state $\left|\Omega\left(t_f\right)\right\rangle$ at $t_f>t_i$. Suppose that we have picked an initial state and electric potential so that in some region of space at the final time $t_f$ the time derivative of the charge expectation of the final state $\left|\Omega\left(t_f\right)\right\rangle$ is non-zero, that is,

$$\frac{\partial\rho_e\left(\mathbf{x},t_f\right)}{\partial t_f}\neq 0 \text{ in some region} \tag{22}$$

How are we justified in assuming that this it is possible to achieve this condition? Well we assume that our theory models the real world. And in the real world there are many situations where the time derivative of the charge density is non-zero. A trivial example would be an electrically charged object moving with any non-zero velocity.

Next consider a similar problem but using, instead, the electric potential,

$$\left(\mathbf{A}',A_0'\right)=\left(\mathbf{A}-\nabla\chi,A_0+\frac{\partial\chi}{\partial t}\right) \tag{23}$$

where $\chi\left(\mathbf{x},t\right)$ is a real valued function that satisfies,

$$\chi\left(\mathbf{x},t\right)=0 \text{ for } t=t_i \text{ and } \left.\frac{\partial\chi\left(\mathbf{x},t\right)}{\partial t}\right|_{t=t_i}=0 \tag{24}$$

Note that at the initial time $t_i$, the quantity $\chi\left(\mathbf{x},t\right)$ and its first derivative with respect to time is zero. Therefore at the initial time the electric potential $\left(\mathbf{A}',A_0'\right)=\left(\mathbf{A},A_0\right)=0$.

Under the action of the electric potential of Eq. (23) the initial state vector $\left|\Omega\left(t_i\right)\right\rangle$ will evolve into the final state vector $\left|\Omega'\left(t_f\right)\right\rangle$ where, in general, $\left|\Omega'\left(t_f\right)\right\rangle\neq\left|\Omega\left(t_f\right)\right\rangle$ because $\left(\mathbf{A}',A_0'\right)$ does not equal $\left(\mathbf{A},A_0\right)$. However we can use (20)

to show that $(\mathbf{A}',A_0')$ is related to $(\mathbf{A},A_0)$ by a gauge transformation. We have postulated that the theory must be gauge invariant. Therefore the current and charge expectation values must be the same for both problems, i.e.,

$$\mathbf{J}_e(\mathbf{x},t) = \langle\Omega(t)|\hat{\mathbf{J}}(\mathbf{x})|\Omega(t)\rangle = \langle\Omega'(t)|\hat{\mathbf{J}}(\mathbf{x})|\Omega'(t)\rangle \tag{25}$$

and,

$$\rho_e(\mathbf{x},t) = \langle\Omega(t)|\hat{\rho}(\mathbf{x})|\Omega(t)\rangle = \langle\Omega'(t)|\hat{\rho}(\mathbf{x})|\Omega'(t)\rangle \tag{26}$$

Also the quantity $\xi(t_i \to t)$ must be the same in both cases because it is also gauge invariant. Use these results in (13) to obtain,

$$\langle\Omega'(t_f)|\hat{H}_0|\Omega'(t_f)\rangle = \xi(t_i \to t) + \int d\mathbf{x}\big(\mathbf{J}_e(\mathbf{x},t_f)\cdot\mathbf{A}'(\mathbf{x},t_f) - \mathbf{J}_e(\mathbf{x},t_i)\cdot\mathbf{A}'(\mathbf{x},t_i)\big)$$
$$- \int_{t_i}^{t_f} dt\int A_0'\left(\frac{\partial\rho_e}{\partial t} + \nabla\cdot\mathbf{J}_e\right)d\mathbf{x} + \langle\Omega(t_i)|\hat{H}_0|\Omega(t_i)\rangle$$

$$\tag{27}$$

Next recall, per Eq. (24), that $\chi(\mathbf{x},t_i) = 0$ and use (23), (24), and (13) in the above to obtain,

$$\langle\Omega'(t_f)|\hat{H}_0|\Omega'(t_f)\rangle = \langle\Omega(t_f)|\hat{H}_0|\Omega(t_f)\rangle - \int d\mathbf{x}\big(\mathbf{J}_e(\mathbf{x},t_f)\cdot\nabla\chi(\mathbf{x},t_f)\big)$$
$$- \int_{t_i}^{t_f} dt\int\frac{\partial\chi}{\partial t}\left(\frac{\partial\rho_e}{\partial t} + \nabla\cdot\mathbf{J}_e\right)d\mathbf{x} \tag{28}$$

Use Eq. (24) and assume reasonable boundary conditions and integrate by parts to obtain,

$$\langle\Omega'(t_f)|\hat{H}_0|\Omega'(t_f)\rangle = \langle\Omega(t_f)|\hat{H}_0|\Omega(t_f)\rangle + \int d\mathbf{x}\chi(\mathbf{x},t_f)\nabla\cdot\mathbf{J}_e(\mathbf{x},t_f)$$
$$- \int\chi(\mathbf{x},t_f)\left(\frac{\partial\rho_e(\mathbf{x},t_f)}{\partial t_f} + \nabla\cdot\mathbf{J}_e(\mathbf{x},t_f)\right)d\mathbf{x} \tag{29}$$
$$+ \int_{t_i}^{t_f} dt\int\chi\frac{\partial}{\partial t}\left(\frac{\partial\rho_e}{\partial t} + \nabla\cdot\mathbf{J}_e\right)d\mathbf{x}$$

This becomes,

$$\left\langle \Omega'\left(t_f\right)\middle|\hat{H}_0\middle|\Omega'\left(t_f\right)\right\rangle = \left\langle \Omega\left(t_f\right)\middle|\hat{H}_0\middle|\Omega\left(t_f\right)\right\rangle - \int \chi\left(\mathbf{x},t_f\right)\left(\frac{\partial \rho_e\left(\mathbf{x},t_f\right)}{\partial t_f}\right)d\mathbf{x}$$
$$+ \int_{t_i}^{t_f} dt \int \chi\left(\mathbf{x},t\right)\frac{\partial L\left(\mathbf{x},t\right)}{\partial t}d\mathbf{x} \tag{30}$$

where,

$$L\left(\mathbf{x},t\right) \equiv \left(\frac{\partial \rho_e\left(\mathbf{x},t\right)}{\partial t} + \nabla\cdot\mathbf{J}_e\left(\mathbf{x},t\right)\right) \tag{31}$$

Now the quantities $\left\langle \Omega\left(t_f\right)\middle|\hat{H}_0\middle|\Omega\left(t_f\right)\right\rangle$, $L\left(\mathbf{x},t\right)$, and $\partial \rho_e\left(\mathbf{x},t_f\right)/\partial t_f$, which appear on the right of Eq. (30), are all independent of $\chi\left(\mathbf{x},t\right)$. That is, $\chi\left(\mathbf{x},t\right)$ can take on any value, subject to the initial conditions (24) without changing the values of the rest of the quantities on the right hand side of (30). We will use this fact to show that we can always find a $\chi\left(\mathbf{x},t\right)$ which makes $\left\langle \Omega'\left(t_f\right)\middle|\hat{H}_0\middle|\Omega'\left(t_f\right)\right\rangle$ a negative number. This is a direct contradiction to the relationship given by (19).

Note that the equation $L\left(\mathbf{x},t\right) = 0$ is the continuity equation. Since local charge conservation is an experimental fact it would be reasonable, at this point, to set $\partial L\left(\mathbf{x},t\right)/\partial t$ to zero. However, we have not assumed local charge conservation, just gauge invariance, therefore we shall consider two possible cases. First, consider the case where $\partial L\left(\mathbf{x},t\right)/\partial t = 0$. In this case Eq. (30) becomes,

$$\left\langle \Omega'\left(t_f\right)\middle|\hat{H}_0\middle|\Omega'\left(t_f\right)\right\rangle = \left\langle \Omega\left(t_f\right)\middle|\hat{H}_0\middle|\Omega\left(t_f\right)\right\rangle - \int \chi\left(\mathbf{x},t_f\right)\left(\frac{\partial \rho_e\left(\mathbf{x},t_f\right)}{\partial t_f}\right)d\mathbf{x} \tag{32}$$

Next set $\chi\left(\mathbf{x},t_f\right) = f\left(\dfrac{\partial\rho_e\left(\mathbf{x},t_f\right)}{\partial t_f}\right)$ where f is a real constant. Use this in the above to obtain,

$$\left\langle\Omega'\left(t_f\right)\middle|\hat{H}_0\middle|\Omega'\left(t_f\right)\right\rangle = \left\langle\Omega\left(t_f\right)\middle|\hat{H}_0\middle|\Omega\left(t_f\right)\right\rangle - f\int\left(\dfrac{\partial\rho_e\left(\mathbf{x},t_f\right)}{\partial t_f}\right)^2 d\mathbf{x} \qquad (33)$$

Given (22) it is evident that the integral is the above expression is greater than zero.

Therefore if f is sufficiently large then $\left\langle\Omega'\left(t_f\right)\middle|\hat{H}_0\middle|\Omega'\left(t_f\right)\right\rangle$ will be negative. Now

consider the case where $\partial L\left(\mathbf{x},t\right)/\partial t$ is non-zero. In this case let,

$$\chi\left(\mathbf{x},t\right) = \left\langle\begin{matrix} -f\partial L\left(\mathbf{x},t\right)/\partial t \text{ for } t_f > t > t_i \\ 0 \text{ for } t = t_f \end{matrix}\right. \qquad (34)$$

Use this in (30) to obtain,

$$\left\langle\Omega'\left(t_f\right)\middle|\hat{H}_0\middle|\Omega'\left(t_f\right)\right\rangle = \left\langle\Omega\left(t_f\right)\middle|\hat{H}_0\middle|\Omega\left(t_f\right)\right\rangle - f\int_{t_i}^{t_f}dt\int\left(\dfrac{\partial L\left(\mathbf{x},t\right)}{\partial t}\right)^2 d\mathbf{x} \qquad (35)$$

As in the previous case if f is sufficiently large than $\left\langle\Omega'\left(t_f\right)\middle|\hat{H}_0\middle|\Omega'\left(t_f\right)\right\rangle$ will be

negative.

What we have shown is that if we assume that the theory is gauge invariant then there must exist state vectors whose free field energy is negative with respect to the vacuum state. This is in contradiction to (19). Therefore if we assume that QED is gauge invariant we find that there is an inconsistency in the formal theory.

## 4. Discussion

Let us review the results so far. We start with equations (1) and (2). These equations govern the dynamics of the theory. From these we then use the basic rules of algebra and calculus to derive (13). Note that the current and charge operators are defined in terms of the field operators in Eq. (4), however this relationship is not used (it is introduced for informational purposes only). Therefore Eq. (13) follows directly from (1) and (2).

So far the theory is incomplete. We have a dynamical equation that describes how a state vector evolves in time but we not have provided any information on the state vectors that the operators act on. In Section 2 we make some comments on the state vectors. The key relationship derived in that section is Eq. (19) which states that the free field energy of any arbitrary state vector must be non-negative.

Note that the theory at this point is still incomplete. To complete the formal theory we would have to define the field operators, in the usual manner, in terms of raising and lowering operators and define the commutator relationships between these operators. The state vectors are then defined in terms of creation operators acting on the vacuum state. At this point we would be able to, in principle, solve actual problems.

However, it is not the purpose of this paper to introduce the complete formal theory which is more then adequately covered in the physics literature. The purpose of this paper is to examine some of the elements of the formal theory and determine if they are mathematically consistent.

In Section 3 we show that these elements are not mathematically consistent. Here we examine an initial state that is acted on by two different electric potentials that differ by a gauge transformation. If we assume that the theory is gauge invariant then we can assume that the physical observables for both potentials are the same. This allows us to derive (30) from (13). From (30) it is easy to show that there must exist state vectors $|\Omega\rangle$ such that $\langle\Omega|\hat{H}_0|\Omega\rangle < 0$ which violate the relationship given by (19).

Note that (13) follows directly from the dynamical equations and (30) follows from (13) and the assumption of gauge invariance. Therefore the fact that there must exist states where $\langle\Omega|\hat{H}_0|\Omega\rangle < 0$ follows from the dynamical equation and the assumption of gauge invariance. Therefore for a mathematically consistent theory the state vector must be defined in such a way that there exist states whose free field energy is less than the vacuum state. A possible way to do this was discussed in [2,4].

## 5. The Vacuum current

At this point we have proved that there is a mathematical inconsistency in some of the elements that make up formal QED. In order to examine this inconsistency further we will solve equation (13) for an actual problem. In order to evaluate (13) for a given applied electromagnetic field we need to be able to determine the current and charge expectation values which is a complicated problem. However, if the initial state is the vacuum state $|0\rangle$ then there are expressions from the literature that give the current and charge expectation value as a function of the applied electromagnetic field for small fields. So in the limit that the applied electromagnetic field is small the we can solve equation (13) if the initial state is $|0\rangle$. In this paper the results of G. Scharf [5] will be used.

Using standard 4-vector notation the Fourier transform of the vacuum current is given in terms of the Fourier transform of the electric potential by (See Section 2.10 of [5] ),

$$J^{\mu}_{vac}(k) = \frac{e^2}{8\pi^2}\left(k^{\mu}k^{\nu} - g^{\mu\nu}k^2\right)\frac{\Pi(k)A_{\nu}(k)}{k^2} \tag{36}$$

where,

$$\Pi(k) = -\frac{2k^4}{3}\int_{4m^2}^{\infty}ds\frac{f(s)}{\left(s - k^2 - ik^0\varepsilon\right)} \tag{37}$$

and

$$f(s) = \frac{\left(s + 2m^2\right)\sqrt{1 - \left(\frac{4m^2}{s}\right)}}{s^2} \tag{38}$$

where $\varepsilon$ is an infinitesimally small positive quantity. From Maxwell's equations,

$$\left(k^{\mu}k^{\nu} - g^{\mu\nu}k^2\right)A_{\nu}(k) = 4\pi J^{\mu}(k) \tag{39}$$

where $J^{\mu}(k)$ is the applied classical 4-current that is the source of the electromagnetic field. Make the substitution $k^0 = \omega$ and use the above result to obtain,

$$J^{\mu}_{vac}(\omega, \mathbf{k}) = (2\pi)^2\eta\,\Pi_1(\omega, \mathbf{k})J^{\mu}(\omega, \mathbf{k}) \tag{40}$$

where $\eta \equiv \dfrac{e^2}{3\pi(2\pi)^2}$ and,

$$\Pi_1(\omega, \mathbf{k}) = \int_{4m^2}^{\infty}\frac{f(s)\left(\omega^2 - |\mathbf{k}|^2\right)}{\left(\omega - \chi_{\mathbf{k}} + i\varepsilon\right)\left(\omega + \chi_{\mathbf{k}} + i\varepsilon\right)}ds \tag{41}$$

where $\chi_{\mathbf{k}} \equiv \sqrt{|\mathbf{k}|^2 + s}$. Take the inverse Fourier transform of (40) to obtain,

$$\mathbf{J}_{vac}(\mathbf{x}, t) = \eta\int\Pi_1(\omega, \mathbf{k})\mathbf{J}(\omega, \mathbf{k})e^{-i\omega t}e^{i\mathbf{k}\cdot\mathbf{x}} \tag{42}$$

Now we will work the following problem. At the initial time $t_i = -\infty$ assume that the system is in the vacuum state $\left|0\right\rangle$ and the applied electric potential is zero. Next apply a non-zero electric potential. According to the Schrödinger equation (1) this will cause the initial vacuum state to evolve into some other state $\left|\Omega\left(t_f\right)\right\rangle$ for $t_f > t_i$. Assume that this applied electrical potential is sufficiently small so that we can use equation (42) for the vacuum current. Using the results for the vacuum current in (13) we can then solve for $\left\langle\Omega\left(t_f\right)\right|\hat{H}_0\left|\Omega\left(t_f\right)\right\rangle$.

Before proceeding we note that it is easy to show that $J_{vac}^\mu$ obeys the continuity equation, i.e.,

$$\left(\frac{\partial\rho_{vac}}{\partial t} + \nabla\cdot\mathbf{J}_{vac}\right) = 0 \tag{43}$$

We can use this in (13) along with the fact that the initial state $\left|\Omega\left(t_i\right)\right\rangle = \left|0\right\rangle$ for this problem so that $\left\langle\Omega\left(t_i\right)\right|\hat{H}_0\left|\Omega\left(t_i\right)\right\rangle = 0$ and the initial vacuum current $\mathbf{J}_{vac}\left(\mathbf{x}, t_i\right) = 0$ to obtain,

$$\left\langle\Omega\left(t_f\right)\right|\hat{H}_0\left|\Omega\left(t_f\right)\right\rangle = \xi\left(t_i \rightarrow t_f\right) + \int d\mathbf{x}\mathbf{J}_{vac}\left(\mathbf{x}, t_f\right)\cdot\mathbf{A}\left(\mathbf{x}, t_f\right) + O\left(A^3\right) \tag{44}$$

where the quantity $O\left(A^3\right)$ means terms to the third order in the electric potential or higher. The rest of the expression on the right of the equals sign is to the second order in the electric potential when we use (42) for the vacuum current. Therefore we will drop the $O\left(A^3\right)$ term and just assume the electric potential is sufficiently small for the equality to hold to the second order.

Now let us apply the above results to the following problem. Assume that the system is in initially in the vacuum state. Apply a classical current which is given by,

$$\mathbf{J}(\mathbf{x},t) = \begin{cases} \mathbf{x}_2 e^{\lambda t} \cos(\beta x_1) \text{ for } t < 0 \\ 0 \text{ for } t \geq 0 \end{cases} \tag{45}$$

where $\mathbf{x}_j$ is a unit vector in the jth direction and $\lambda > 0$. We will evaluate $\langle \Omega(t_f) | \hat{H}_0 | \Omega(t_f) \rangle$ at $t_f = \varepsilon$ where $\varepsilon \to 0^+$, that is, $\varepsilon$ approaches zero from above so that $\varepsilon$ is an infinitesimal positive quantity. In order to evaluate $\langle \Omega(\varepsilon) | \hat{H}_0 | \Omega(\varepsilon) \rangle$ we refer back to (44) and evaluate the quantities on the right of the equal sign for $t_f = \varepsilon$. The calculations are straightforward but somewhat lengthy therefore they have been done the Appendix. The key results are that $\xi(-\infty \to \varepsilon) < 0$ and $\int \mathbf{J}_{vac}(\mathbf{x},\varepsilon) \cdot \mathbf{A}(\mathbf{x},\varepsilon) d\mathbf{x} < 0$. We use this in (44) to obtain,

$$\langle \Omega(\varepsilon) | \hat{H}_0 | \Omega(\varepsilon) \rangle < 0 \tag{46}$$

However this contradicts the relationship given in Section 2, $\langle \Omega | \hat{H}_0 | \Omega \rangle \geq 0$ for all possible state vectors $|\Omega\rangle$. This is in agreement with the results of Section 3 and confirms that there is a mathematical inconsistency in the formal theory.

## 6. Conclusion.

We have examined some of the elements of the formal theory of QED. These were the Schrödinger equation introduced in Section 1, a requirement that the free field energy of the state vectors is non-negative as discussed in Section 2, and the assumption of gauge invariance as discussed in Section 3. It was shown that these elements are not mathematically consistent. In addition we solved Eq. (13) for the case where the initial state is the vacuum state. We did this by using results from the literature for the vacuum current. This allowed us to solve for the vacuum current, electric field, and vector potential for the case when the applied current given by Eq. (45). This confirms the results of the previous section. It is evident that for the formal theory of QED to be mathematically consistent then there must exist state vectors whose free field energy is negative with respect to the vacuum state.

## **Appendix**

We want to solve (44) for the case where the applied current is given by Eq. (45) and the initial state is the vacuum state. This allows us to use (42) for the vacuum current. First take the Fourier transform of (45) to obtain,

$$\mathbf{J}(\omega,\mathbf{k}) = \int \mathbf{J}(\mathbf{x},t)e^{i(\omega t-\mathbf{k}\cdot\mathbf{x})}d\omega d\mathbf{k} = \left(\frac{-i\mathbf{x}_2}{2(\omega-i\lambda)}\right)\left(\delta(\beta-k_1)+\delta(\beta+k_1)\right)\delta(k_2)\delta(k_3)$$

$$(47)$$

Use this in (42) to obtain,

$$\mathbf{J}_{vac}(\mathbf{x},t) = -i\eta\mathbf{x}_2 \int\limits_{4m^2}^{\infty} f(s)ds \int d\omega d\mathbf{k} \frac{\delta(k_2)\delta(k_3)\left(\omega^2-|\mathbf{k}|^2\right)e^{-i(\omega t-\mathbf{k}\cdot\mathbf{x})}}{(\omega-\chi_{\mathbf{k}}+i\epsilon)(\omega+\chi_{\mathbf{k}}+i\epsilon)}\left(\frac{\left(\delta(\beta-k_1)+\delta(\beta+k_1)\right)}{2(\omega-i\lambda)}\right)$$

$$(48)$$

This yields,

$$\mathbf{J}_{vac}(\mathbf{x},t) = -i\eta\mathbf{x}_2 \int\limits_{4m^2}^{\infty} f(s)ds \int d\omega \frac{\left(\omega^2-\beta^2\right)e^{-i\omega t}}{(\omega-\chi_{\beta}+i\epsilon)(\omega+\chi_{\beta}+i\epsilon)}\left(\frac{\cos(\beta x_1)}{(\omega-i\lambda)}\right) \quad (49)$$

For $t < 0$ we obtain,

$$\mathbf{J}_{vac}(\mathbf{x},t<0) = 2\pi\eta \int\limits_{4m^2}^{\infty} f(s) \frac{\left(\lambda^2+\beta^2\right)\cos(\beta x_1)e^{\lambda t}}{\left(\lambda^2+\beta^2+s\right)}ds \quad (50)$$

For $t \geq 0$

$$\mathbf{J}_{vac}\left(\mathbf{x}, t \geq 0\right) = -2\pi\eta\mathbf{x}_2 \int\limits_{4m^2}^{\infty} \frac{sf(s)}{2\chi_\beta}\left(\frac{e^{-i\chi_\beta t}}{\left(\chi_\beta - i\lambda\right)} + \frac{e^{+i\chi_\beta t}}{\left(\chi_\beta + i\lambda\right)}\right)\cos(\beta x_1)ds \tag{51}$$

This becomes,

$$\mathbf{J}_{vac}\left(\mathbf{x}, t \geq 0\right) = -2\pi\eta\mathbf{x}_2\cos(\beta x_1)\int\limits_{4m^2}^{\infty}\frac{sf(s)}{\chi_\beta}\left(\frac{\chi_\beta\cos\left(\chi_\beta t\right) + \lambda\sin\left(\chi_\beta t\right)}{\beta^2 + \lambda^2 + s}\right)ds \tag{52}$$

To solve for the electromagnetic field we will work in the Lorentz gauge so that,

$$\frac{\partial A_0}{\partial t} + \nabla \cdot \mathbf{A} = 0 \tag{53}$$

This yields,

$$\frac{\partial^2 \mathbf{A}}{\partial t^2} - \nabla^2 \mathbf{A} = 4\pi\mathbf{J} \tag{54}$$

From this and (45) it is easy to show that,

$$\mathbf{A}\left(\mathbf{x}, t\right) = \begin{cases} \dfrac{\mathbf{x}_2 4\pi e^{\lambda t}\cos\left(\beta x_1\right)}{\lambda^2 + \beta^2} \text{ for } t < 0 \\ \dfrac{\mathbf{x}_2 4\pi\cos\left(\beta x_1\right)}{\beta\left(\lambda^2 + \beta^2\right)}\left(\beta\cos\left(\beta t\right) + \lambda\sin\left(\beta t\right)\right) \text{ for } t \geq 0 \end{cases} \tag{55}$$

where we have used the fact that $\mathbf{A}\left(\mathbf{x}, t\right)$ and its first derivative with respect to time is continuous at $t = 0$. From the above we obtain,

$$\mathbf{E}\left(\mathbf{x}, t\right) = -\frac{\partial \mathbf{A}\left(\mathbf{x}, t\right)}{\partial t} = \begin{cases} -\dfrac{\mathbf{x}_2 4\pi\lambda e^{\lambda t}\cos\left(\beta x_1\right)}{\lambda^2 + \beta^2} \text{ for } t < 0 \\ \dfrac{\mathbf{x}_2 4\pi\cos\left(\beta x_1\right)}{\left(\lambda^2 + \beta^2\right)}\left(\beta\sin\left(\beta t\right) - \lambda\sin\left(\beta t\right)\right) \text{ for } t \geq 0 \end{cases} \tag{56}$$

Use (56) and (50) to obtain,

$$\xi(-\infty \to 0) = \int\limits_{-\infty}^{0} dt \int \mathbf{J}_{vac}(\mathbf{x},t) \cdot \mathbf{E}(\mathbf{x},t) d\mathbf{x} = \left(-4\pi^2\eta\right)\left(\frac{V}{2}\right) \int\limits_{4m^2}^{\infty} \frac{f(s)ds}{\left(\lambda^2+\beta^2+s\right)} \tag{57}$$

where V is the integration volume and we have used $\int\left(\cos(\beta x_1)\right)^2 d\mathbf{x} = V/2$. Next calculate,

$$\xi(0 \to t) = \int\limits_{0}^{t} dt' \int \mathbf{J}_{vac}(\mathbf{x},t') \cdot \mathbf{E}(\mathbf{x},t') d\mathbf{x} \tag{58}$$

Use (52) and (56) in the above to obtain,

$$\xi(0 \to t) = \left(\frac{-8\pi^2\eta}{\left(\lambda^2+\beta^2\right)}\right)\left(\frac{V}{2}\right) \int\limits_{4m^2}^{\infty} \frac{sf(s)ds}{\chi_\beta\left(\lambda^2+\beta^2+s\right)}\{D_1+D_2+D_3+D_4\} \tag{59}$$

where,

$$D_1 = -\beta\chi_\beta\left(\frac{\cos\left[(\beta+\chi_\beta)t\right]-1}{2(\beta+\chi_\beta)} - \frac{\cos\left[(\chi_\beta-\beta)t\right]-1}{2(\chi_\beta-\beta)}\right)$$

$$D_2 = -\lambda\chi_\beta\left(\frac{\sin\left[(\beta+\chi_\beta)t\right]}{2(\beta+\chi_\beta)} + \frac{\sin\left[(\chi_\beta-\beta)t\right]}{2(\chi_\beta-\beta)}\right)$$

$$D_3 = -\lambda\beta\left(\frac{\sin\left[(\beta+\chi_\beta)t\right]}{2(\beta+\chi_\beta)} - \frac{\sin\left[(\chi_\beta-\beta)t\right]}{2(\chi_\beta-\beta)}\right)$$

$$D_4 = \lambda^2\left(\frac{\cos\left[(\beta+\chi_\beta)t\right]-1}{2(\beta+\chi_\beta)} + \frac{\cos\left[(\chi_\beta-\beta)t\right]-1}{2(\chi_\beta-\beta)}\right)$$

Next calculate $\int \mathbf{J}_{vac} \cdot \mathbf{A} d\mathbf{x}$ at $t \geq 0$. Use (55) and (52) to obtain,

$$\int \mathbf{J}_{vac} \cdot \mathbf{A} d\mathbf{x} \underset{t \geq 0}{=} \left(-8\pi^2\eta\right)\left(\frac{V}{2}\right)\left\{ \left[\left[\int\limits_{4m^2}^{\infty}\left(\frac{sf(s)\left(\chi_\beta\cos\left(\chi_\beta t\right)+\lambda\sin\left(\chi_\beta t\right)\right)}{\chi_\beta\left(\beta^2+\lambda^2+s\right)}\right)\right]\right] \times\left(\frac{\beta\cos\left(\beta t\right)+\lambda\sin\left(\beta t\right)}{\beta\left(\lambda^2+\beta^2\right)}\right) \right\} \tag{60}$$

This yields,

$$\int \mathbf{J}_{vac} \cdot \mathbf{A} d\mathbf{x} \underset{t \geq 0}{=} \left(-8\pi^2\eta\right)\left(\frac{V}{2}\right)\left[\int\limits_{4m^2}^{\infty}\left(\frac{sf(s)}{\beta\chi_\beta\left(\beta^2+\lambda^2\right)\left(\beta^2+\lambda^2+s\right)}\right)\right]\left\{\begin{array}{l}\beta\chi_\beta\cos\left(\chi_\beta t\right)\cos\left(\beta t\right)\\+\beta\lambda\sin\left(\chi_\beta t\right)\cos\left(\beta t\right)\\+\lambda\chi_\beta\cos\left(\chi_\beta t\right)\sin\left(\beta t\right)\\+\lambda^2\sin\left(\chi_\beta t\right)\sin\left(\beta t\right)\end{array}\right\}ds \tag{61}$$

We choose to evaluate $\left\langle\Omega(t)\left|\hat{H}_0\right|\Omega(t)\right\rangle$ at $t=\varepsilon$ where $\varepsilon\to 0^+$, that is, $\varepsilon$ approaches zero from above so that epsilon is an infinitesimal positive quantity. This point is chosen due to the fact that it easy to evaluate. In order to evaluate $\left\langle\Omega(\varepsilon)\left|\hat{H}_0\right|\Omega(\varepsilon)\right\rangle$ we refer back to (44) and evaluate the quantities on the right of the equal sign for $t=\varepsilon$. We obtain,

$$\xi\left(-\infty\to\varepsilon\right)=\xi\left(-\infty\to 0\right)+\xi\left(0\to\varepsilon\right) \tag{62}$$

On examining (59) we can show that $\xi\left(0\to\varepsilon\right)\underset{\varepsilon\to 0^+}{=}0$ therefore,

$$\xi\left(-\infty\to\varepsilon\right)\underset{\varepsilon\to 0^+}{=}\xi\left(-\infty\to 0\right)<0 \tag{63}$$

In addition, at $t=\varepsilon$,

$$\int \mathbf{J}_{vac} \cdot \mathbf{A} d\mathbf{x} \underset{t=\varepsilon}{=} \left(-8\pi^2\eta\right)\left(\frac{V}{2}\right)\left[\int\limits_{4m^2}^{\infty}\left(\frac{sf(s)\cos\left(\chi_\beta\varepsilon\right)}{\left(\beta^2+\lambda^2\right)\left(\beta^2+\lambda^2+s\right)}\right)\right]ds<0 \tag{64}$$

Therefore both terms on the right side of the Eq. (44) are negative. The result is that at

$t = \varepsilon$, $\left\langle \Omega(\varepsilon) \middle| \hat{H}_0 \middle| \Omega(\varepsilon) \right\rangle < 0$.